\documentclass[apj]{emulateapj}

\usepackage{apjfonts}

\newcommand{\iso}[2]{\hbox{${}^{#1}{\rm #2}$}}
\newcommand{\Msun}{\ensuremath{{M}_{\sun}}}

\shorttitle{The $^{18}$F($\alpha$,$p$) reaction and AGB stars}
\shortauthors{Karakas et al.}

\begin{document}


\title{The impact of the \iso{18}F($\alpha$,$p$)\iso{21}Ne reaction 
on asymptotic giant branch nucleosynthesis}


\author{Amanda I. Karakas\altaffilmark{1,2,3}}
\affil{Research School of Astronomy \& Astrophysics, Mt Stromlo Observatory,
Weston Creek ACT 2611, Australia}
\email{akarakas@mso.anu.edu.au}

\author{Hye Young Lee\altaffilmark{1,4}}
\affil{Physics Division, Argonne National Laboratory, Argonne, IL 60439-4843}
\email{hylee@phy.anl.gov}

\author{Maria Lugaro\altaffilmark{1,5}}
\affil{Sterrenkundig Instituut, University of Utrecht, Postbus 80000, 3508 TA Utrecht, The Netherlands}
\email{M.Lugaro@phys.uu.nl}

\author{J. G\"{o}rres and M. Wiescher}
\affil{Department of Physics and Joint Institute for Nuclear Astrophysics, 
University of Notre Dame, Notre Dame, IN 46556}
\email{jgorres@nd.edu,mwiesche@nd.edu}


\altaffiltext{1}{First three authors contributed equally to this work.}
\altaffiltext{2}{Visiting Scholar, Physics Division, Argonne National Laboratory, Argonne, IL 60439-4843}
\altaffiltext{3}{Department of Astronomy and Astrophysics, 
University of Chicago, 5640 S. Ellis Avenue, Chicago, Illinois 60637}
\altaffiltext{4}{Department of Physics and Joint Institute for Nuclear Astrophysics, 
University of Notre Dame, Notre Dame, IN 46556}
\altaffiltext{5}{Centre for Stellar \& Planetary Astrophysics, Monash University,
Clayton VIC 3800, Australia}


\begin{abstract}
We present detailed models of low and intermediate-mass asymptotic
giant branch (AGB) stars with and without the
$^{18}$F($\alpha$,$p$)$^{21}$Ne reaction included in the nuclear
network, where the rate for this reaction has been recently
experimentally evaluated for the first time. 
The lower and recommended measured rates for this
reaction produce negligible changes to the stellar yields,
whereas the upper limit of the rate affects the production of 
$^{19}$F and $^{21}$Ne. The stellar yields increase
by $\sim 50$\% to up to a factor of 4.5 for $^{19}$F, and  
by factors of $\sim 2$ to 9.6 for $^{21}$Ne.
While the $^{18}$F($\alpha$,$p$)$^{21}$Ne 
reaction competes with $^{18}$O production, the extra protons 
released are captured by $^{18}$O to facilitate the 
$^{18}$O($p$,$\alpha$)$^{15}$N($\alpha,\gamma$)$^{19}$F chain.
The higher abundances of $^{19}$F obtained using the upper limit of the 
rate helps to match the [F/O] ratios observed in AGB stars, but only for 
large C/O ratios. Extra-mixing processes are proposed to help
to solve this problem. Some evidence that the
$^{18}$F($\alpha$,$p$)$^{21}$Ne rate might be closer to
its upper limit is provided by the fact that the higher calculated
$^{21}$Ne/$^{22}$Ne ratios in the He intershell provide an 
explanation for the Ne isotopic composition of silicon-carbide 
grains from AGB stars. 
This needs to be confirmed by future experiments of the 
$^{18}$F($\alpha$,$p$)$^{21}$Ne reaction rate. The availability of 
accurate fluorine yields from AGB stars will be fundamental for
interpreting observations of this element in carbon-enhanced 
metal-poor stars.
\end{abstract}




\section{Introduction}

Interest in the $^{18}$F($\alpha$,p)$^{21}$Ne reaction
($Q$ value=1.741 MeV) came from early pre-supernova (SN) models
that suggested that the reaction might be important in the helium
and carbon burning regions during the SN. After the shock wave 
increases the internal temperature and density, the 
timescale for destruction of $^{18}$F via the ($\alpha$,p) 
reaction is comparable to that of its $\beta^{+}$-decay lifetime 
\citep{arnett69,truran78,ugiesen87}, where the laboratory half-life of
\iso{18}F is $\tau_{1/2} = 109$ minutes. The early work by 
\citet{arnett69} used unpublished theoretical estimates from Fowler;
these rates did not appear in \citet{fowler75}, \citet{harris83}
nor \citet{cf88}, and are only valid for $T \ge 800 \times 10^{6}$K. 
Until 2006 the only rate for the 
$^{18}$F($\alpha$,p)$^{21}$Ne reaction was 
the theoretical estimate available in the Brussels nuclear
reaction-rate library \citep{aikawa05}.  The first experiment 
aimed at determining the $^{18}$F($\alpha$,p)$^{21}$Ne rate 
over a large range of stellar temperatures was
carried out by Lee et al. (2007, in preparation). This experimental 
evaluation, when considering its associated uncertainties,
presented significant differences compared to the 
theoretical rate, especially at the low temperatures relevant for
He-shell burning in AGB stars ($T \approx 300 \times 10^{6}$K).
In this paper we investigate the effect of such differences 
on the nucleosynthesis occurring in AGB models of various
initial mass and composition.

These are stars of mass less than $\sim 8\Msun$ 
located in the high-luminosity and low-temperature region of 
the Hertzsprung-Russell diagram. They have evolved through 
core H and He burning, and are now sustained against gravitational 
collapse by alternate H and He-shell burning 
\citep[see][for a recent review]{herwig05}.  
AGB stars are the site of nucleosynthesis and mixing processes 
that lead to the production of carbon, nitrogen, fluorine and heavy 
elements such as barium and lead. The strong stellar winds associated 
with these stars ensure that the freshly synthesized material is 
expelled into the interstellar medium, making AGB stars major 
factories for the production of the elements in the Universe 
\citep{busso99}.

The theoretical estimate of the $^{18}$F($\alpha$,p)$^{21}$Ne rate
was not present in our previous works \citep{lugaro04,karakas06a}, 
although we had included the
species \iso{18}F because of its important role in the reaction
chain \iso{14}N($\alpha,\gamma$)\iso{18}F($\beta^+\nu$)\iso{18}O
leading to the production of \iso{18}O in the He shell. 
In this note we include this reaction in the network and study 
its effects in detail because preliminary results showed
an enhanced production of \iso{19}F when employing
the new upper limit of the $^{18}$F($\alpha$,p)$^{21}$Ne rate.
This is of interest because AGB
models do not synthesize enough \iso{19}F to match the [F/O]
abundances observed in AGB stars \citep{jorissen92,forestini92}.
This negative result remains even after examining most of the 
current error bars of the many reactions involved in the
complex chain of production of \iso{19}F in AGB
stars, such as the \iso{14}C($\alpha,\gamma$)\iso{18}O and the
\iso{19}F($\alpha,p$)\iso{22}Ne reactions \citep{lugaro04}. There
are still uncertainties in the stellar models that could affect
the match to the observations, in particular extra-mixing
processes, as proposed by \citet{lugaro04}. However, we will not
be able to accurately pin down the effects of such uncertain stellar
processes while our estimates of the abundance of \iso{19}F in
AGB stars are still undermined by uncertainties in the reaction
rates involved. 

The cosmic origin of fluorine is not yet completely understood.
Type II SN explosions \citep{woosley95} and stellar winds from 
Wolf Rayet stars \citep{meynet00} both play a significant role 
in producing this fragile element alongside AGB stars 
\citep{renda04}. Observationally, AGB stars and their progeny 
(e.g. post-AGB stars, planetary nebulae) are the only confirmed 
site of fluorine production
thus far \citep{jorissen92,werner05,zhang05,pandey06}, with
no clear indication for enhanced F abundances resulting from the 
$\nu$-process in a region shaped by past SNe 
\citep{federman05}.  Moreover, the recent observations of a
greatly enhanced F abundance ([F/Fe] = 2.90) in a Carbon-Enhanced
Metal-Poor (CEMP) halo star polluted via mass transfer from a
companion during its AGB phase \citep{schuler07}
represents further strong motivation to better understand the
details of the fluorine production mechanism in AGB stars.
 
The $^{18}$F($\alpha$,p)$^{21}$Ne reaction could also affect the 
abundance of \iso{21}Ne in the He-shell of AGB stars. 
There is a long-standing puzzle concerning
the isotopic composition of Ne measured in stellar silicon carbide
(SiC) grains extracted from meteorites, which formed in the
extended envelopes of carbon-rich AGB stars. 
About 40\% of these grains contain $^{22}$Ne and/or $^{4}$He of 
nucleosynthetic origin \citep{heck07}. Being a noble gas, Ne
is believed to be ionized and implanted in the SiC dust during the
very last phases of AGB evolution \citep{lewis94,verchovsky04}.
Measurements performed on
a large number of grains show that the observed Ne composition can
be explained by the mixing of He-shell matter into the envelope
material of AGB stars \citep{lewis90,gallino90,lewis94,heck07}.
While the $^{20}$Ne/$^{22}$Ne ratios are well
reproduced in this scenario\footnote{The extreme enrichment of
$^{22}$Ne in these materials is historically known as the Ne-E(H)
component in meteorites, whose presence was one of the keys leading
to the discovery of stellar SiC grains in meteorites \citep{anders93}.},
the $^{21}$Ne/$^{22}$Ne ratios are higher than
predicted by AGB models. \citet{lewis94} attributed the higher
than predicted abundance of $^{21}$Ne to spallation reactions where
the grains are bombarded by cosmic rays during their residence
time in the interstellar medium. These authors hence related the
excesses of $^{21}$Ne with respect to the values predicted by AGB
models to the age of the grains. However, \citet{ott00}
have shown experimentally that the majority of
pre-solar SiC grains would have essentially lost all the $^{21}$Ne
produced during spallation  by recoil. These authors suggest that
the observed variations of the $^{21}$Ne/$^{22}$Ne ratios in SiC
grains are more likely due to the effect of nucleosynthesis in the
He-burning shell of their parent AGB star, and this is also
indicated by their correlation with nucleosynthesis effects in the
Kr isotopic ratios. The identification of such nucleosynthesis
effects, however, are to date missing. The 
$^{18}$F($\alpha$,p)$^{21}$Ne reaction could play a role 
in this puzzle.

For these reasons we aim to explore in detail the effect of
the new experimental evaluation of the
$^{18}$F($\alpha$,p)$^{21}$Ne rate, briefly described in \S\ref{rate}, 
on the production of fluorine and $^{21}$Ne in detailed AGB models.
Our methods and models are presented in \S\ref{models}, results
in \S\ref{results} and \S\ref{ne21}, and we finish with a 
discussion and conclusions.

\section{The \iso{18}F($\alpha,p$)\iso{21}Ne reaction rate} \label{rate}

\begin{figure}
\begin{center}
\plotone{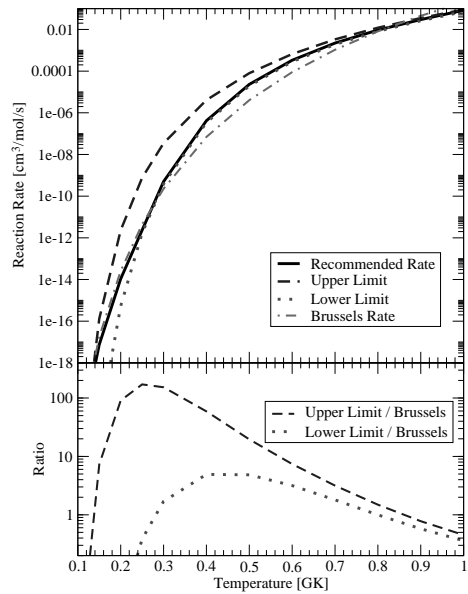}
\caption{Reaction rate of $^{18}$F($\alpha$,p)$^{21}$Ne including 
the upper and lower limits. Also shown is the Brussels theoretical
estimate of this rate. In the lower panel, the ratios of the current 
upper and lower limits with respect to the Brussels rate are shown.
\label{fig:reac-rate}}    
\end{center}  
\end{figure}

The measurement of the $^{18}$F($\alpha$,p)$^{21}$Ne reaction 
cross section is made difficult by the 
short half-life of $^{18}$F. Owing to the problems associated
with the production of a long-lived 
\iso{18}F target or a high intensity \iso{18}F beam, the first study 
of this reaction was based on the measurement of the time-reversed 
$^{21}$Ne(p,$\alpha$)$^{18}$F reaction at the Dynamitron Tandem 
Laboratory Bochum \citep{ugiesen87}. 
The cross section measurements at higher proton energies 
($E_{p} > 3$~MeV) were based on the direct spectroscopy of the 
emitted $\alpha$ particles, while the lower energy range was 
investigated using the activation method by analyzing 
the decay of \iso{18}F. The results were not published because the low 
energy data were affected by the strong beam-induced background from 
the \iso{18}O$(p,n)$\iso{18}F reaction.  With the development of an 
intense \iso{18}F beam at the Cyclotron Research Center at the 
Universit\'{e} de Louvain la Neuve, a direct measurement of the 
$^{18}$F($\alpha$,p)$^{21}$Ne reaction became possible and the 
reaction cross section was determined at higher energies 
\citep[$E_{\alpha} > 1.4$~MeV corresponding to $E_{p} > 3.1$~MeV,][]{lee06}.
The beam intensity, however, was not sufficient to extend these 
first measurements to energies of relevance for stellar He burning. 
In a complementary experiment therefore, the low energy range
($E_{\rm p} \le 2.3$ MeV) of $^{21}$Ne(p,$\alpha$)$^{18}$F was re-investigated 
at the 4~MV Van de Graaff accelerator at the University of Notre Dame 
using the activation method 
\citep[][Lee et al. 2007, in preparation]{LeeThesis}.

The cross section of the \iso{18}O$(p,n)$\iso{18}F
background reaction was measured independently over the entire 
energy range. The corresponding \iso{18}F activity was normalized 
to the abundance of \iso{18}O impurities in the target, and 
subtracted from the \iso{21}Ne$(p,\alpha)$ induced \iso{18}F activity. 
Based on these data, a reaction rate for  $^{18}$F($\alpha$,p)$^{21}$Ne
was determined for the stellar temperature range 
0.2 $\le T$~(GK)~$\le 1.0$ of relevance for AGB star nucleosynthesis. 
The lower limit of the cross-section measurement is mainly determined 
by the statistical uncertainty of the activation data, while the 
upper limit is based on the uncertainty associated with the 
\iso{18}O induced background. The resulting cross-section 
data were analyzed in terms of the R-matrix theory. 
The analysis, however, was hampered by the lack of detailed information 
about the specific parameters of the observed resonances. 
Fig.~\ref{fig:reac-rate} shows the reaction rate as a function of 
temperature based on these recent 
measurements. The solid black line indicates the recommended rate while 
the long-dashed line and the dotted lines show the upper and lower limits, 
respectively. These limits of the reaction rate correlate with the
experimental uncertainties in the cross-section data, as well as 
with the uncertainties from nuclear structure information.
Shown for comparison is the predicted Hauser-Feshbach 
rate as a gray dot \citep{aikawa05}. The present recommended 
rate is in good agreement with the Hauser-Feshbach prediction in 
the characteristic temperature range of AGB stars. Details of the 
experiment and the reaction rate analysis will be discussed 
in a forthcoming paper (Lee et al. 2007, in preparation).

\section{The Stellar Models} \label{models} 

The numerical method we use has been previously described in detail
\citep{lugaro04,karakas06a}. Here we summarize the main points relevant 
for this study.   We computed the stellar structure first using 
the Mt Stromlo Stellar Structure code \citep{lattanzio86}, and then 
performed post-processing on that structure to obtain abundances 
for 77 species, most of which are not included in the small 
stellar-structure network.
This technique is valid for studying reactions not directly related 
to the main energy generation, as they can be assumed to have no impact 
on the stellar structure
This is certainly the case for studying the effect of the 
\iso{18}F($\alpha,p$)\iso{21}Ne reaction on AGB nucleosynthesis. 
On top of including neutron-capture reaction rates from 
\citet{bao00} for nuclei from Ne to S, the main change to the 
nuclear network for this study is the addition of the 
\iso{18}F($\alpha,p$)\iso{21}Ne reaction rate into the 
77 species network.  

The stellar-structure models used for this study are summarized
in Table~\ref{tab:models}, and have been previously discussed in
detail in \citet[][and references therein]{karakas07b}.  Owing to 
the fact that we found the \iso{18}F($\alpha,p$)\iso{21}Ne reaction 
to affect the abundance of \iso{19}F we have concentrated on models 
that produce the most of it i.e. $M \sim 3\Msun$ \citep{lugaro04}.
We also show results from a lower mass (1.9$\Msun$) and two
intermediate-mass (5$\Msun$) AGB stars for comparison. Both
the 5$\Msun$ models experience proton-capture nucleosynthesis
at the base of the convective envelope (hot bottom burning, HBB).
The 3$\Msun$, $Z=0.012$ model was computed with the revised solar
abundances from \citet{asplund05}, whereas the $Z=0.02$ models 
were computed with \citet{anders89} abundances. The
lower metallicity models were computed using \citet{anders89}
scaled-solar abundances. We also present a model for a
2$\Msun$ $Z=$0.0001 ([Fe/H] $\sim -2.3$) star, which is relevant 
to the above-mentioned recent observation of highly-enhanced 
fluorine in a halo star of similar metallicity.

A partial mixing zone (PMZ) is required to produce a \iso{13}C 
{\it pocket} in the He-intershell during the interpulse period. 
It is in the \iso{13}C pocket that neutrons are released by the 
\iso{13}C($\alpha, n$)\iso{16}O reaction \citep{gallino98}; 
in this study we artificially include a PMZ of constant mass 
at the deepest extent of each third dredge-up (TDU) mixing 
episode in exactly the same
way as described by \citet{lugaro04}. We include a pocket of 
0.002$\Msun$ for all lower mass cases, and 
we include a pocket of $1\times 10^{-4}\Msun$ into the 
5$\Msun$, $Z=0.02$ model.  Note that these choices result
in a \iso{13}C pocket 
between 10\% to 15\% of the mass of the He-intershell region.

In Table~\ref{tab:models} we present the initial mass and 
metallicity, $Z$, the C, N and O solar abundances used in the structure model
where AG89 refers to \citet{anders89} and A05 to \citet{asplund05},
the mass of the partial mixing zone (PMZ),
the total number of thermal pulses (TPs) computed, the maximum 
temperature in the He-shell, $T_{\rm He}^{\rm max}$, the maximum 
temperature at the base of the convective envelope, $T_{\rm bce}^{\rm max}$, 
the total mass mixed into the envelope by TDU episodes, Mass$_{\rm dred}$, 
and the final envelope mass $M_{\rm env}$.  
All data are in solar units, except the temperatures, which
are in millions of kelvins.  We present some information about 
the light elements including the surface C/O and \iso{12}C/\iso{13}C 
number ratios at the last computed time step.

\section{Results} \label{results} 

In Table~\ref{tab:yields} we show results from the stellar models 
that employed the recommended rate of the \iso{18}F($\alpha,p$)\iso{21}Ne 
reaction. For each mass and $Z$ value, we show the C, N and O abundances
used in the structure model (as for Table~\ref{tab:models}), the mass of 
the PMZ used in the computation, the yield ($y$) of \iso{19}F, the 
production factor ($f$) of \iso{19}F, and the multiplication factor 
($X$) needed to obtain the upper limit \iso{19}F yield from the 
recommended-rate yield.
All yields are in solar masses, the production factors $f$ and the
multiplication factors are dimensionless quantities.
The same information is also presented for \iso{21}Ne for each model.
We compute stellar yields by integrating the surface abundances lost 
in the wind over the stellar lifetime, normalized to the initial abundance 
in the wind \citep[see for e.g.][]{karakas07b}.  The production 
factors are defined according to $f = \log_{10} (X_{\rm end} / X_{\rm initial})$,
where $X_{\rm end}$ is the mass fraction at the tip of the AGB and 
$X_{\rm initial}$ is the initial mass fraction.  The yields from the 
recommended calculations are essentially the same as the yields obtained
from models that employed the lower limit, adopted the Brussels 
theoretical rate, or did not include the \iso{18}F($\alpha,p$) 
reaction at all.

From inspection of Table~\ref{tab:yields} 
we can see that employing the new upper limit of the 
\iso{18}F($\alpha,p$)\iso{21}Ne reaction results in a significant
increase in the production of \iso{19}F and \iso{21}Ne.
The change in the yield increases with decreasing metallicity, 
at a given mass, with the largest change found in the 
5$\Msun$, $Z=0.004$  model where the \iso{19}F yield increased 
by a factor of  4.5. The largest change in the \iso{21}Ne
yield is a factor of 9.6 for the 3$\Msun$, $Z = 0.008$ model.
While we find large increases in the F yield for both the 
intermediate-mass AGB models, the absolute yields are significantly
smaller than those from the lower mass objects; 
this is because \iso{19}F is destroyed by HBB.
For example the 5$\Msun$, $Z = 0.02$ model produced 3 times less
\iso{19}F than the 3$\Msun$, $Z=0.02$ case, whereas the 
5$\Msun$, $Z = 0.004$ model produced about 40 times less \iso{19}F
than the 2.5$\Msun$, $Z=0.004$ model.
From  Table~\ref{tab:yields} we note that the PMZ had little
effect on the production of \iso{19}F and \iso{21}Ne in the
5$\Msun$, $Z=0.02$ model.

While increases in the \iso{21}Ne yield as a consequence 
of using the upper limit of the
\iso{18}F($\alpha,p$) rate are larger than for \iso{19}F, the 
overall amount of this isotope produced by AGB stars remains small.
This is reflected in the production factors that are 
$f \lesssim 0.3$ for all models but the 2.5$\Msun$, $Z=0.004$ 
and 2$\Msun$, $Z=0.0001$ models, where the production factors
are 0.45~dex and 2.08~dex, respectively.  The increase at
very low metallicity might be significant for chemical
evolution studies of the Ne isotopes. Overall however,
we conclude that the rare isotope \iso{21}Ne is not significantly 
produced in AGB stars, even when using the upper limit of the 
\iso{18}F($\alpha,p$) reaction in the calculations.
Most of this isotope in the Galaxy originates from Type 
II SN \citep{woosley95,timmes95}, although it would still
be an interesting exercise to include our AGB yields 
into a chemical evolution model.  The impact of the upper 
limit on \iso{21}Ne production is more important for 
stellar SiC grains, this is discussed further in 
\S\ref{ne21}.

In this section we did not discuss the surprising result
that the \iso{18}F($\alpha, p$)\iso{21}Ne reaction affects the
production of \iso{19}F in AGB stars. It is not intuitive why
this should be the case, so in the next section we outline the
mechanism responsible for the production of the extra fluorine.

\subsection{The \iso{19}F production mechanism}

The enhanced abundance of $^{19}$F may be explained by considering
the $^{18}$O($p, \alpha$)$^{15}$N($\alpha,\gamma)^{19}$F reaction
chain. Including the $^{18}$F($\alpha, p$)$^{21}$Ne reaction
reduces the abundance of $^{18}$O because it competes with
$^{18}$O production via the $^{18}$F($\beta^+\nu)^{18}$O decay.
However, the extra amount of protons from ($\alpha, p$) enhances
the $^{18}$O($p, \alpha$)$^{15}$N reaction rate, even though
$^{18}$O production has been deprived from the decay. In other
words, the sum $N_{^{18}{\rm O}}+N_{p}$
(where $N_{i}$ is the abundance by number of nucleus $i$) remains
constant, however, the product $N_{^{18}{\rm O}} N_{p}$, on
which the number of $^{18}$O+$p$ reactions depends, is maximized
when $N_{^{18}{\rm O}}$ is equal to $N_{p}$.

We can analytically analyze the effect of the extra protons 
on the $^{19}$F production in the He-shell. We simplify the 
$^{18}$O($p, \alpha$)$^{15}$N($\alpha,\gamma)^{19}$F 
reaction chain to the
$^{18}$O($p, \alpha$)$^{15}$N reaction. 
In a He-rich region,  all $^{14}$N is converted to 
$^{18}$F via the ($\alpha,\gamma)$ reaction; this 
either decays to $^{18}$O via the $\beta^+$-decay with a 
branching ratio of $f$ or makes extra protons via the 
($\alpha, p$) reaction with a branching ratio of $1-f$.  Then, 
the number density of $^{18}$O, $N_{^{18}{\rm O}}$, is written
as $f N_{^{14}{\rm N}}$, and for protons, $N_{p}$, as 
$N_{{p}_0} + (1-f)  N_{^{14}{\rm N}}$, where 
$N_{{p}_0}$ is the original number density of protons
without the inclusion of the \iso{18}F($\alpha,p$)\iso{21}Ne
reaction, and $N_{^{14}{\rm N}}$ is the \iso{14}N from the 
H-burning ashes.
Then, the reaction rate of $^{18}$O($p, \alpha$)$^{15}$N can be 
written as
\begin{eqnarray}
N_{^{18}{\rm O}} N_{p} \langle \sigma v \rangle_{(p,\alpha)} 
 & = & f N_{^{14}{\rm N}} [N_{{p}_0} + (1-f)  N_{^{14}{\rm N}}] \langle \sigma v \rangle_{(p,\alpha)} \\
 & = & fN_{^{14}{\rm N}} N_{{p}_0}  \langle \sigma v \rangle_{(p,\alpha)} 
   \times [1+(1-f)N_{^{14}{\rm N}}/N_{{p}_0}] \label{18Op}
\end{eqnarray}
Since $fN_{^{14}{\rm N}} N_{{p}_0}\langle \sigma v \rangle_{(p,\alpha)}$ is 
the rate of $^{18}$O(p,$\alpha$)$^{15}$N without including the 
$^{18}$F($\alpha$,p) reaction, the term [$1+(1-f)N_{^{14}{\rm N}}/N_{{p}_0}$] 
may be thought of as an ``$^{19}$F enhancement factor''.  
The overall \iso{19}F production increases as long as 
$N_{^{14}{\rm N}}/N_{{p}_0} > 1$, and this condition is well 
satisfied in the He-burning shell. During the network calculation a 
realistic $N_{^{14}{\rm N}}/N_{{p}_0} \approx 10^{10}$; 
this ratio is large enough to explain the enhanced fluorine 
production in the stellar models.

As possible sources of uncertainty we can ignore the other 
\iso{18}F $+ \alpha$ channels, that is the ($\alpha,n$) and the 
($\alpha, \gamma$). According to the Brussels theoretical estimate
\citep{aikawa05} the ($\alpha, \gamma$) reaction is approximately 
two orders of magnitude slower at 0.3GK than the ($\alpha,p$), 
whereas the ($\alpha,n$) is 40 orders of magnitude slower.

\section{\iso{21}Ne in meteoritic SiC grains} \label{ne21}

\begin{figure}
\begin{center}
\plotone{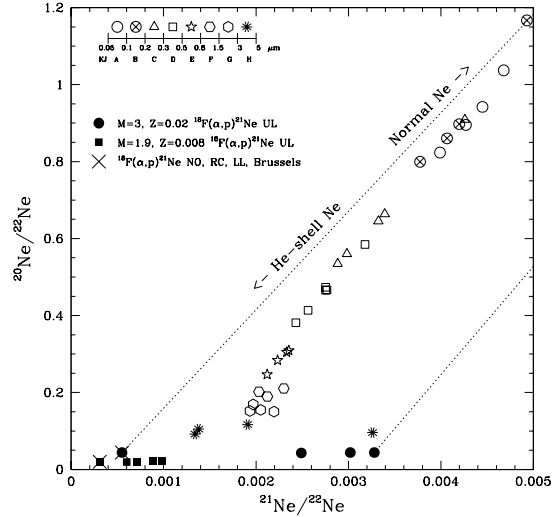}
\caption{Ne isotopic ratios observed in meteoritic SiC 
grains and predicted in the intershell of our 3, $Z=$0.02 
and 1.9$\Msun$, $Z=$0.008 models. The plot is a reproduction of 
Fig. 8 of \citet{lewis94}, where we have added the model predictions. 
For each model we plot the Ne isotopic ratios in the He intershell 
at the end of each TP occurring when C/O$>$1 is satisfied 
in the envelope of the star. The crossed full symbols represent 
models computed without the $^{18}$F($\alpha$,p)$^{21}$Ne reaction
rate, which give a constant result. The full symbols represent models
run using the upper limit of the $^{18}$F($\alpha$,p)$^{21}$Ne
reaction rate. Dotted lines connect the ``Normal Ne'' component of 
solar composition to the ``He-shell Ne'' component corresponding to 
the final compositions of the intershell for the 
3$\Msun$, $Z=$0.02 model. 
\label{fig:ne21}}    
\end{center}  
\end{figure}

To address the puzzle of the $^{21}$Ne/$^{22}$Ne ratio in stellar
SiC grains we have analyzed the effect of using the new
$^{18}$F($\alpha$,p)$^{21}$Ne reaction rate on the $^{21}$Ne
abundance in the He intershell of AGB stars. The results are shown
in Fig.~\ref{fig:ne21} and compared to the SiC data from
\citet{lewis94}.  The plot is a
reproduction of Fig.~8 of \citet{lewis94} where we have added
our new model predictions. The SiC data are derived from measurements 
on samples of grains {\it in bulk}, i.e. collections of a large
number ($\sim$millions) of grains. Different symbols represent
measurements done on collections of grains sampling different
sizes, from 0.01 to 5 $\mu$m, as described in the figure. 
Note that, since measurements in bulk are performed on millions of
grains, they can be only used to derive the average properties of
the parent stars of the grains.

Each data point in Fig.~\ref{fig:ne21} is interpreted as having
been produced by a mixture between the material initially present
in the envelope of the star and the material mixed from the He
intershell into the envelope by TDU. These two ``ingredients'' are
referred in the plot as the ``Normal Ne'' and the ``He-shell Ne''
components, respectively. The Normal component is taken to have 
solar composition. The SiC grains show a composition dominated by
a ``He-shell'' component extremely enhanced in $^{22}$Ne with
respect to solar, as it is the composition of the He
intershell of AGB stars. However, it is clear that the data points
do not lie on the straight mixing line between the two components
(the dotted lines in the plot), which means that the ``He-shell''
component must be variable if we want to account for all the
different measurements.

Model predictions presented in the plot are for the 3$\Msun$
$Z=$0.02 and 1.9$\Msun$ $Z=$0.008 models. These models are the
best within our sample listed in Table~\ref{tab:models} to 
represent the parent stars of SiC grains. This is because they 
reach carbon-rich conditions toward the end of their evolution 
(a necessary condition for the formation of SiC) and have masses 
(between 1.5 and 3$\Msun$), and metallicities (close to solar)
 in the range of the best candidate SiC parent star models 
\citep[see e.g.][for a thorough discussion]{lugaro99,lugaro03b}. 

When we compute our models using the recommended, lower limit, or
Brussels theoretical evaluation of the
$^{18}$F($\alpha, p$)$^{21}$Ne reaction rate, the results are
equivalent to the models computed without the inclusion of this
reaction, and they are the same as those presented by
\citet{gallino90}. The $^{21}$Ne/$^{22}$Ne ratio in the intershell
is constant $\simeq$ 0.0004 and the rightward shift to higher
$^{21}$Ne/$^{22}$Ne ratios observed in the grains cannot be
reproduced. Note that in this case the abundances of $^{20}$Ne and
$^{21}$Ne are barely modified in the intershell, in particular
$^{21}$Ne is destroyed by factors 5 to 50 in the H-burning ashes
and restored to its original Solar System value by neutron-capture 
reactions on $^{20}$Ne during the TPs, with neutrons 
released by the $^{22}$Ne($\alpha, n$)$^{25}$Mg reaction.
One the other hand, models computed with the upper limit of the
$^{18}$F($\alpha, p$)$^{21}$Ne reaction rate show an increase in
the $^{21}$Ne abundance, and hence in the $^{21}$Ne/$^{22}$Ne
ratio in the intershell of up to a factor of 6, which is the
number needed to reach up to the most extreme data point observed
at $^{21}$Ne/$^{22}$Ne=0.0033. The predicted intershell
$^{21}$Ne/$^{22}$Ne ratio increases with pulse number and with
the stellar mass because the temperature increases and the
$^{18}$F($\alpha$,p)$^{21}$Ne reaction becomes more efficient.
The last computed TPs reached 302 and 278 $\times 10^{6}$K for 
the 3 $\Msun$ and the 1.9 $\Msun$ models, respectively.

Another possible way of producing a higher abundance of $^{21}$Ne
in the He intershell is by increasing the neutron-capture cross
section of $^{20}$Ne. The value we use is 0.199 mbarn at 30 keV,
which is recommended by \citet{bao00} and corresponds to the
experimental estimate of \citet{winters88}. A much higher value of
1.5 mbarn at 30 keV was previously suggested by \citet{almeida83},
in which case the final $^{21}$Ne/$^{22}$Ne ratio in the intershell
of our 3$\Msun$ $Z=$0.02 model is equal to 0.002. 
However, the data of \citet{almeida83}
have recently been re-analyzed (M. Heil, personal communication)
resulting in a much lower cross section of 0.303 mbarn at 30 keV.
With this latest evaluation the final $^{21}$Ne/$^{22}$Ne ratio in
the intershell of our 3$\Msun$ $Z=$0.02 model reaches only 0.00073.
We also checked that possible changes in the neutron capture cross 
section of $^{21}$Ne itself, and the current uncertainties of the 
\iso{22}Ne($\alpha,n$)\iso{25}Mg reaction rate \citep{karakas06a} 
do not lead to significant variations in the abundance of this isotope. 
These considerations lead us to conclude that the 
\iso{18}F($\alpha,p$)\iso{21}Ne reaction rate being close to its 
upper limit would be a promising explanation for the $^{21}$Ne/$^{22}$Ne 
ratios in SiC grains.

Finally, we note that increasing $^{21}$Ne/$^{22}$Ne ratios are  
correlated with increasing $^{86}$Kr/$^{82}$Kr ratios measured in 
SiC grains \citep[see Fig.5 of][]{ott00}. This correlation can be 
qualitatively matched by considering that both the $^{21}$Ne and 
$^{86}$Kr abundances in the intershell increase with increasing 
temperature. This is because $^{86}$Kr is produced via the branching 
point at $^{85}$Kr during the high-neutron density flux produced by 
the the $^{22}$Ne($\alpha, n$)$^{25}$Mg reaction during TPs 
\citep[see e.g.][]{abia01}. Quantitatively, however, our 
models can only match the lowest observed $^{86}$Kr/$^{82}$Kr. 
It remains to be seen if this mis-match can be attributed to 
uncertainties in the nuclear properties of the $^{85}$Kr branching 
point, or to intershell temperatures higher than those of our 
models during the late AGB or the post-AGB phases.
Further work is needed to address this point.

\section{Discussion} 

\begin{figure}
\begin{center}
\plotone{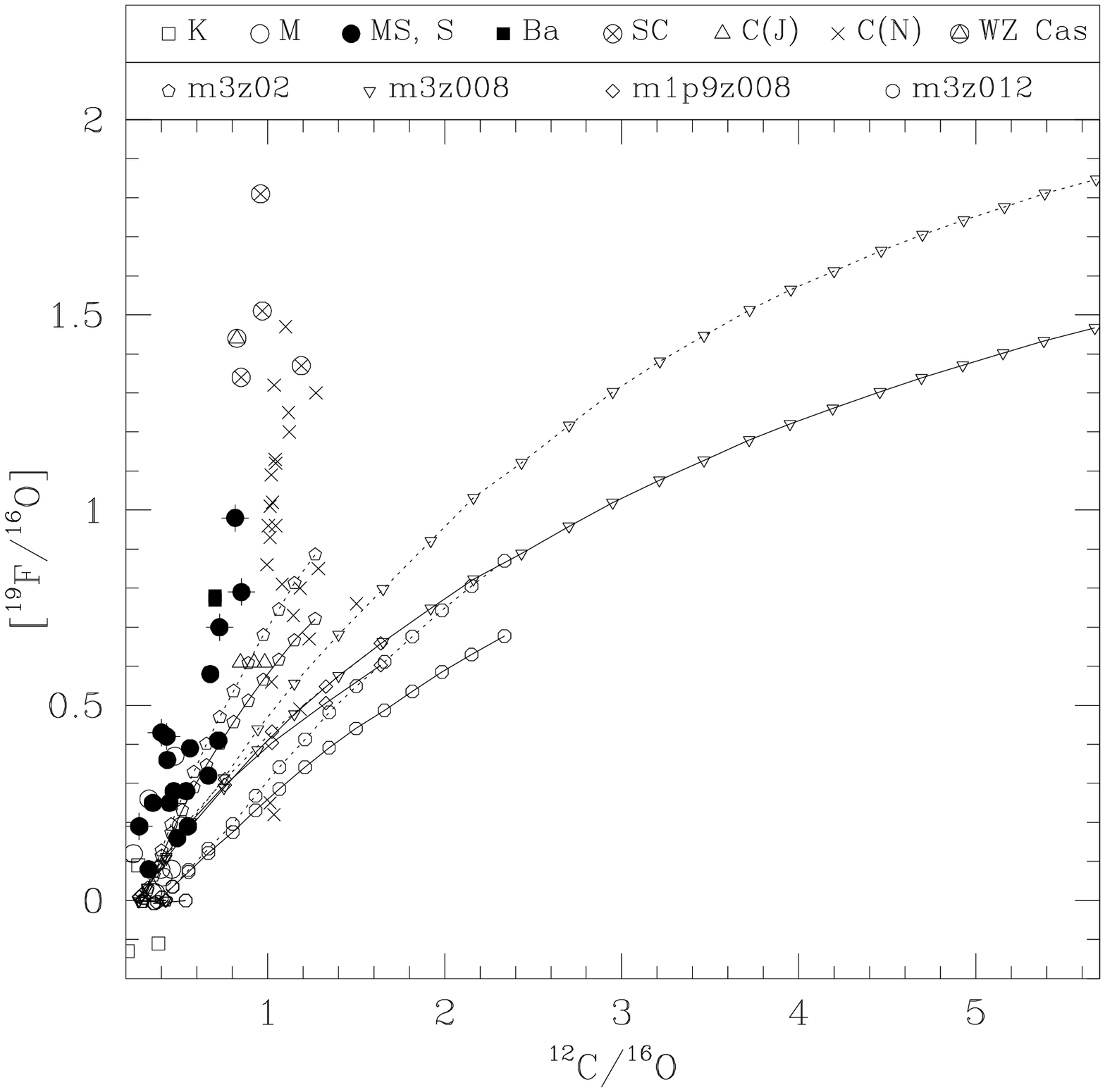}
\caption{Comparison of fluorine abundances observed by \citet{jorissen92}
 and model predictions for selected stellar models: 
3$\Msun$ with $Z$=0.02, 0.012, and 0.008; and 1.9$\Msun$ with $Z$=0.008.
All models include a PMZ of $0.002\Msun$.
Predictions are normalized in such way that the initial \iso{19}F abundance 
corresponds to the average F abundance observed in K and M stars 
\citep[see][]{jorissen92}. Crossed MS and S symbols denote stars with 
large N excesses. Each symbol on the prediction lines 
represents a TDU episode. Solid lines represent calculations performed 
using no \iso{18}F($\alpha,p$)\iso{21}Ne reaction, which are equivalent 
to using the current lower limit, recommended value and Brussels library 
rate. Dotted lines are calculations performed using the current upper 
limit of the rate. 
\label{fig:f19stars}}  
\end{center}
\end{figure}

In Fig.~\ref{fig:f19stars} we show the evolution of the 
surface [\iso{19}F/\iso{16}O] ratio as function of the 
C/O ratio for four AGB models, compared to the
observations of fluorine-enhanced stars from \citet{jorissen92}. 
The models are selected to best represent the features of the 
observed stars, that is, stars with masses in the range 
1.5 to 3$\Msun$ \citep{wallerstein98}, and with metallicities 
around $Z=0.01$. Similarly to the yields, the final surface 
[\iso{19}F/\iso{16}O] ratios from these models are roughly 50 
to $\sim$ 140\% higher when calculations are done using the 
new upper limit of the \iso{18}F($\alpha,p$)\iso{21}Ne reaction.

From Fig.~\ref{fig:f19stars} we see that using the new upper
limit of the \iso{18}F($\alpha,p$)\iso{21}Ne reaction can result
in a match between the stellar models and the stars with
the highest observed \iso{19}F abundances, but only for the very
high C/O ratios of $\sim 4-5$, found in the 3$\Msun$, $Z = 0.008$ model.
In the lower mass models and in the 3$\Msun$ of solar metallicity,
the new upper limit does not result in a match between
the predicted and observed [F/O] abundances. 
\citet{lugaro04} suggested that extra-mixing processes in AGB
stars may help to solve this problem by converting C into N, hence
decreasing the C/O ratio for a given \iso{19}F abundance.
Further indication of this possibility is the
fact that for any given C/O ratio MS and S stars with the higher
\iso{19}F abundance also have N excesses, and lower
\iso{12}C/\iso{13}C ratios than predicted by standard models
\citep{abia97}.
Detailed studies of the possible effects of extra-mixing phenomena
are required, and will have to analyze the impact of using the
higher \iso{19}F abundance obtained using the upper limit of the
\iso{18}F($\alpha,p$)\iso{21}Ne reaction rate.

There are many uncertainties that affect AGB
stellar models including the treatment of convection
and mass loss \citep[see][for a detailed discussion]{herwig05}.
One modeling uncertainty that might affect the results
is that most of the stellar models did not lose all of their
convective envelopes when the evolution sequences ended, that
is, they did not leave the AGB track, and could, in principle,
experience extra TPs and TDU episodes.  This possibility is 
discussed in \citet{karakas07a},  where it was estimated that 
one more TP may occur for e.g. the 3$\Msun$, $Z=0.012$ model.
We do not repeat this exercise here owing to the
uncertainty of the efficiency of the TDU at small
envelope masses \citep[see discussion in][]{karakas07b}, 
but note that more TDU episodes would further enrich the 
\iso{19}F and \iso{12}C abundances at the stellar surface. 

Another modeling uncertainty that will affect our results 
is the choice of mass-loss rate during the AGB. 
We used the \citet{vw93} mass-loss prescription that was 
empirically derived from Mira-type variables and might overestimate 
mass loss for semiregulars, thus terminating the TP-AGB phase too 
early and hampering the formation of C stars at low masses. 
At solar metallicity we do not form carbon-rich stars with
initial masses below 2.5$\Msun$ \citep{karakas02}, 
whereas typical C-star {\em initial} 
masses are $\sim 2\Msun$ \citep{claussen87}, although 
this result is somewhat model dependent \citep{abia01,kahane00}. 
Regardless, this observational result is in contradiction 
to our models, and is caused partly by our choice of mass loss, 
and also because we do not find efficient enough (or any) 
TDU in the low-mass AGB models of $\approx Z_{\odot}$.  
Certainly, a different choice of mass loss would have a 
significant effect on the stellar structure and on the 
resulting F and \iso{21}Ne yields. We address this point 
in \citet{karakas06a} for intermediate-mass AGB stars 
where the yields of \iso{25}Mg and \iso{26}Mg changed
by more than an order of magnitude by using the Reimers
mass-loss rate on the AGB instead of \citet{vw93}; we
speculate that we would expect similar changes to the 
yields of lower mass stars but more work is needed
to address this important point.
One final comment is that the 3$\Msun$, $Z=0.02$ model 
becomes a C-rich star at a total (current) mass of 
$\sim 2.3 \Msun$. Given the uncertainties in deriving
total masses of C stars this is not entirely 
out of the range of expected C-star masses.

\begin{figure}
\begin{center}
\includegraphics[width=7cm,angle=270]{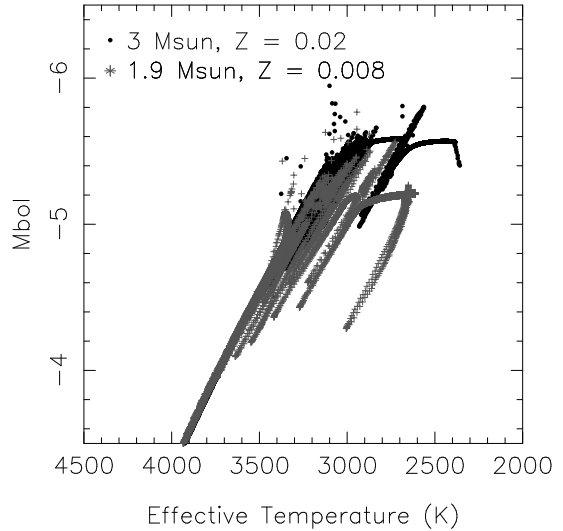}
\caption{Bolometric luminosity, $M_{\rm bol}$, versus effective 
temperature, $T_{\rm eff}$, for the 3$\Msun$, $Z=0.02$ (black dots),
and the 1.9$\Msun$, $Z=0.008$ (gray crosses) models during the TP-AGB phase. 
The large variation in $M_{\rm bol}$ and $T_{\rm eff}$ shown in this 
diagram is caused by the change in these observables during the AGB 
lifecycle (i.e. thermal pulse -- dredge-up -- interpulse).
\label{bolometric}}
\end{center}
\end{figure}

In Fig.~\ref{bolometric} we show the bolometric luminosities plotted
against effective temperature for two stellar models that
become C-rich near the tip of the TP-AGB phase.
Fig.~\ref{bolometric} can be compared to Fig.~6 in \citet{busso07},
with luminosities and temperatures from a selection of 
AGB stellar models computed with the FRANEC code 
\citep{straniero03}, plotted against bolometric luminosities derived
from observations of C-rich stars \citep[see also][]{guandalini06,whitelock06}.
In comparison to the FRANEC models, AGB models computed 
with the Monash stellar structure code cover a similar range of 
$T_{\rm eff}$ from 3,200~K to 2,500~K as most of the carbon stars, 
and cover the observed range of bolometric luminosities. 
Similar to the FRANEC models, we cannot match the $T_{\rm eff}$'s 
of the coolest stars with temperatures $\sim 2000$~K.  
However, we must be
cautious about making conclusions from this comparison
because we are showing the entire AGB evolutionary sequence,
not just the sequences when the model stars have C/O $> 1$
at the surface. Secondly, we do not include a realistic
treatment of low-temperature molecular opacities but instead
we approximated the opacity from CN, CO, ${\rm H}_{2}{\rm O}$ 
and TiO using the formulations prescribed by \citet{bessell89},
and corrected by \citet{chiosi93}.  These are fits to the molecular
opacities of \citet{alexander75} and \citet{alexander83},
and while they do include a dependence on envelope composition
do not treat properly treat C-rich compositions 
\citep[see, for example,][]{marigo02}. 

Both \citet{marigo02} and \citet{busso07} have outlined 
the importance of using realistic low-temperature 
molecular opacities in detailed AGB models. 
Future work will study the effect of carbon-rich
molecular opacities on the stellar structure and
nucleosynthesis. \citet{marigo02} found that the inclusion
of C-rich opacities truncated the TP-AGB evolution 
fairly quickly (that is, in a couple of TPs) once the 
C/O ratio exceeded unity. This is because the C-rich 
molecules that form under such conditions caused the star 
to become larger and cooler, and this in turn increased 
the mass-loss rate.  
One of us (Karakas, Wood \& Campbell, in preparation) is
currently studying the effect of such opacities on detailed
AGB models and noticed similar trends, in that the evolution
ends before the C/O ratio exceeds values much larger than
$\sim 2$. One consequence of this is that we would no
longer predict the large C/O ratios found in the
3$\Msun$, $Z=0.008$ model (see Fig~\ref{fig:f19stars}),
and, subsequently, values of [F/O] greater than $\simeq$1.

Another exciting future opportunity is represented by the 
comparison of our models of very low metallicity, e.g. 
$Z=$0.0001 ([Fe/H] $\sim -2.3$), to observations of fluorine 
in CEMP stars, which likely achieved their chemical 
peculiarities from an AGB companion. \citet{schuler07} observed
[F/Fe] $= 2.90$ in one such star with a [Fe/H] $= -2.5$. The 
2$\Msun$ $Z=$0.0001 model reached a fluorine production factor 
of $\sim 3.59$, which translates into a huge [F/Fe] $=3.63$. 
This may be enough to explain the observations of 
\citet{schuler07}, although we need to consider dilution 
due to binary mass transfer. The fact that fluorine production 
shows a strong dependence on the initial stellar mass 
\citep[see, for example, Fig.~1 of][]{lugaro04}, suggests that
we may use the detailed model predictions along with observations 
of the lowest metallicity stars, to provide constraints on
the properties of the initial mass function in the early Universe
\citep[see, for example,][]{tumlinson07}. 

\section{Conclusions}

In conclusions, the comparison of our results to observations
of [F/O] in AGB stars, and to the Ne composition of SiC grains
suggests that the values of the \iso{18}F($\alpha,p$)\iso{21}Ne 
reaction rate may lie closer to the current upper limit.
More experimental data for this reaction at temperatures 
below 0.4~GK are, however, required to help verify this result.  
The result for F in AGB stars is less compelling than the
results for Ne in SiC grains, owing to the fact that we cannot
match the whole observed [F/O] range. To add to this 
problem is the need for some extra-mixing process to alter the
C and N abundances while not destroying \iso{19}F.
Also, AGB modeling uncertainties (e.g. mass loss and molecular 
opacities) could dramatically affect the predictions 
of F yields and surface abundances, rendering any conclusions
uncertain. 

The modeling uncertainties related to extra-mixing, the TDU and
mass loss do not affect, however, the intershell
compositions of our stellar models and thus do not apply to
the discussion of the Ne composition of stellar SiC grains.
From Fig.~\ref{fig:ne21} and the related discussion, we see that
the measured Ne isotopic compositions could be explained by
the upper limit for the \iso{18}F($\alpha,p$)\iso{21}Ne 
reaction. This tantalizing result is also a more reliable hint
that the reaction is indeed closer to its upper limit than
the comparison to F in AGB stars. However, further work is 
required to test this scenario, including a detailed investigation
into Kr nucleosynthesis in AGB stars. 

Finally, the larger stellar yields of \iso{19}F obtained 
using the upper limit of the \iso{18}F($\alpha,p$)\iso{21}Ne 
reaction should be tested in a galactic chemical evolution model 
of the type presented by \citet{renda04}. An AGB contribution to 
the production of \iso{21}Ne may also be considered, given 
that the upper limit of the \iso{18}F($\alpha,p$) rate
results in a larger production of this rare Ne isotope.
The observations of low F abundances in stars in the
globular cluster $\omega$ Centauri by \citet{cunha03}, where
other observations clearly indicate pollution by AGB stars
\citep[e.g.][]{stanford07}, are puzzling. Clearly further
work is required to address the nucleosynthetic origin of this
most interesting and fragile element.



\acknowledgments

We thank Michael Heil for providing unpublished neutron-capture cross 
section data, Tim Beers for discussions and the referee for a thorough
report that has helped to improve the manuscript.
AIK wishes to thank Ken Nollett and Jim Truran for the opportunity
to spend three months in Chicago, where this paper was written, and
acknowledges partial support from the Joint Theory Institute funded 
together by Argonne National Laboratory and the University of Chicago.
AIK also acknowledges support from the Australian Research Council's 
Discovery Projects funding scheme (project number DP0664105). 
HYL, JG, and MW acknowledge support from the National Science Foundation 
under Grant No. PHY01-40324, the Joint Institute for Nuclear Astrophysics, 
NSF-PFC, under Grant No. PHY02-16783.
ML is supported by the NWO through a VENI fellowship, and wishes to 
thank MW for the hospitality at the University of Notre Dame during 
the time this paper was written.

\bibliography{apj-jour,/home/akarakas/biblio/library}

\clearpage

\begin{table}[t]
\begin{center}
\caption{Data and results from the stellar models, see the text in \S\ref{models}
for details.\label{tab:models}}
\vspace{1mm}
\begin{tabular}{@{}ccccccccccc@{}} 
\tableline\tableline
 Mass &  $Z$ & CNO\tablenotemark{a} & PMZ &  TPs & $T_{\rm He}^{\rm max}$ &  $T_{\rm bce}^{\rm max}$ 
   & Mass$_{\rm dred}$ & $M_{\rm env}$ & C/O & \iso{12}C/\iso{13}C \\
\tableline
 3.0  &  0.02  & AG89 & 0.002 &  26 & 302 & 6.75 & 8.1($-2$) & 0.676  & 1.40 & 118 \\
 5.0  &  0.02  & AG89 & 0     &  24 & 352 & 64.5 & 5.0($-2$) & 1.500  & 0.77 & 7.84  \\ 
 5.0  &  0.02  & AG89 & 1E$-$4 & -- & --  & --   & --        &  --    &  --  &  --  \\
 3.0  &  0.012 & A05  & 0.002 &  22 & 307 & 7.23 & 9.2($-2$) & 0.806  & 2.47 & 168  \\
 1.9  &  0.008 & AG89 & 0.002 &  17 & 278 & 3.29 & 2.2($-2$) & 0.222  & 1.30 & 138  \\ 
 3.0  &  0.008 & AG89 & 0.002 &  29 & 319 & 10.5 & 2.1($-1$) & 0.549  & 5.00 & 519  \\ 
 2.5  &  0.004 & AG89 & 0.002 &  28 & 308 & 7.33 & 1.9($-1$) & 0.685  & 11.9 & 1300  \\
 5.0  &  0.004 & AG89 & 0     &  81 & 377 & 84.4 & 2.2($-1$) & 1.141  & 2.64 & 11.0  \\
 2.0  & 0.0001 & AG89 & 0.002 &  26 & 307 & 9.00 & 2.2($-1$) & 0.040  & 105  & 2.25($+$4) \\
\tableline
\end{tabular}

\tablenotetext{a}{Initial CNO abundances where ``AG89'' refers to \citet{anders89}
initial solar or scaled solar abundances, and ``A05'' refers to \citet{asplund05} 
solar abundances.}

\end{center}
\end{table}

\begin{table}[t]
\begin{center}
\caption{Stellar yields of \iso{19}F and \iso{21}Ne from the AGB models.  
\label{tab:yields}}
\vspace{1mm}
\begin{tabular}{@{}cccccccccc@{}} 
\tableline\tableline
 Mass &  $Z$ & CNO & PMZ & $y$(\iso{19}F$_{\rm rec}$) & $f$(\iso{19}F) & $X$(\iso{19}F) &
 $y$(\iso{21}Ne$_{\rm rec}$) & $f$(\iso{21}Ne) & $X$(\iso{21}Ne) \\
\tableline
 3.0  &  0.02  & AG89 & 0.002  & 5.84($-6$) & 0.684 & 1.526 & 1.25($-6$) & 0.053 & 4.423 \\
 5.0  &  0.02  & AG89 & 0      & 1.83($-6$) & 0.223 & 1.632 & 2.09($-6$) & 0.050 & 3.463 \\
 5.0  &  0.02  & AG89 & 1E$-$4 & 1.87($-6$) & 0.227 & 1.625 & 2.16($-6$) & 0.052 & 3.327 \\
 3.0  &  0.012 & A05  & 0.002  & 5.66($-6$) & 0.676 & 1.736 & 1.39($-6$) & 0.102 & 5.330 \\
 1.9  &  0.008 & AG89 & 0.002  & 9.35($-7$) & 0.583 & 1.178 & 1.60($-7$) & 0.032 & 2.340 \\
 3.0  &  0.008 & AG89 & 0.002  & 1.71($-5$) & 1.466 & 2.407 & 4.52($-6$) & 0.340 & 9.609 \\
 2.5  &  0.004 & AG89 & 0.002  & 1.33($-5$) & 1.752 & 2.061 & 2.81($-6$) & 0.456 & 8.364 \\
 5.0  &  0.004 & AG89 & 0      & 1.45($-7$) & 0.104 & 4.582 & $-$2.58($-6$) & $-$0.627 & $-$1.965 \\
 2.0  & 0.0001 & AG89 & 0.002  & 1.67($-5$) & 3.589 & 1.975 & 3.23($-6$) & 2.080 & 8.551 \\   \tableline
\end{tabular}

\end{center}
\end{table}

\end{document}